\documentstyle[twocolumn,aps,epsfig,floats]{revtex}
\draft 

\begin{document} 
\twocolumn[\hsize\textwidth\columnwidth\hsize\csname @twocolumnfalse\endcsname
\title{
Origin of the giant magnetic moments 
of Fe impurities on and in Cs films} 
\author{S. K. Kwon and B. I. Min} 
\address{Department of Physics, 
         Pohang University of Science and Technology, 
         Pohang 790-784, Korea} 

\date{\today}
 
\maketitle 
 
\begin{abstract}    
To explore the origin of the observed giant magnetic moments 
($\sim 7 \mu_B$) of Fe impurities on the surface and in the bulk of 
Cs films, we have performed the relativistic LSDA + $U$ calculations 
using the linearized muffin-tin orbital (LMTO) band method.  
We have found that Fe impurities in Cs behave differently from 
those in noble metals or in Pd.
Whereas the induced spin polarization of Cs atoms is negligible,
the Fe ion itself is found to be the source of the giant magnetic moment.
The $3d$ electrons of Fe in Cs are localized 
as the $4f$ electrons in rare-earth ions so that
the orbital magnetic moment becomes as large as the spin magnetic moment.   
The calculated total magnetic moment of $M = 6.43 \mu_B$, 
which comes mainly from Fe ion, 
is close to the experimentally observed value. 
\end{abstract}

\pacs{PACS number: 75.20.Hr,71.70.Ej,72.15.Rn}
]
\narrowtext  
The problem of magnetic impurities in metals has been 
an important branch of solid state physics\cite{Hewson}.  
The Anderson impurity Hamiltonian or the Kondo impurity Hamiltonian 
has been a starting point of the model calculations. 
In the side of electronic structure calculation, 
it is difficult to apply the local spin density approximation 
(LSDA)~\cite{Kohn}
in its original form, 
because the Coulomb correlation energy of $3d$ electrons is 
underestimated in the LSDA.

Recently, giant magnetic moments are observed 
in the anomalous Hall effect measurements for the Fe and Co  
impurities on the surface and in the bulk of Cs films\cite{Beckmann}.
The deduced values of magnetic moments are $\sim 7 \mu_B$ and $\sim 8 \mu_B$ 
for Fe and Co impurities, respectively. 
These are unexpectedly large magnetic moments, as compared to $2.22 \mu_B$ 
and $1.72 \mu_B$ for pure Fe and Co metals.   
These features were attributed 
to an induced spin polarization of the neighboring Cs atoms, as in 
the case of $3d$ magnetic impurities in Pd films\cite{Bergmann}.
Pd is a near-magnetic transition element, and so
Pd atoms nearby magnetic Fe impurity are easily spin polarized to yield
giant magnetic moments.
Whether this is also true for Fe impurities in Cs films is a question 
to be addressed in this work.

The magnetic moments of $3d$ transition metal elements 
in solids are usually determined by the spin magnetic moments.
The orbital magnetic moments are quenched 
by the crystal field and/or the hybridization effects. 
In contrast, $4f$ rare-earth  elements
have comparable sizes of the orbital magnetic moments to those of 
the spin magnetic moments. The total magnetic moment $M$ is given by
$M = g_{J}\mu_{B}\left[J(J+1)\right]^{1/2}$,
where $J$ is the total angular momentum determined by the Hund's rule:
$J = |L-S|$ ($L+S$) for $n \leq (2l+1)$ ($n \geq (2l+1)$), 
and $g_{J}$ is the Land$\acute{\mbox{e}}$'s $g$-factor.
According to the formula, a free Fe ion in an electron configuration of
Fe$^{2+}$ will have effectively $M = 6.70 \mu_B$, unless the orbital 
magnetic moment is quenched\cite{Ashcroft}.  
This theoretical magnetic moment is very close to the experimentally 
observed $M=7 \mu_B$ of Fe impurities in Cs~\cite{Beckmann}. 
Therefore, if Fe ions in Cs are isolated because of the large lattice
constant of Cs metal, the main contribution to the observed magnetic 
moments will come from Fe ions themselves rather than from the induced 
spin polarization of the surrounding Cs atoms.  
In fact, previous local susceptibility data of Fe ions 
in heavy alkali-metals have been discussed in line with 
isolated $3d$ states for Fe ions\cite{Riegel1,Riegel2}.  
The present works are to determine what is realized in these systems. 

We have simulated Fe impurities in Cs films 
using the linearized muffin-tin orbital (LMTO) band method\cite{Andersen}.
In our calculations, we assume that Fe impurities are substitutional 
and the ideal body-centered-cubic (bcc) structure of Cs 
($a = 6.05 $ \AA) is maintained even after impurities are introduced. 
Although the assumptions of the substitutional Fe impurities and 
the rigid lattice are not valid {\it a priori}, these effects are thought 
to be not so critical\cite{Gross,Papanikolaou}. 
The function of the basis set is adopted up to $l = 2$ orbitals 
for all the atomic sites.

Since the Coulomb correlation interaction of Fe $3d$ electrons is 
expected to be significant in these systems, the LSDA + $U$ method will 
be useful.
The LSDA + $U$~\cite{Anisimov} removes the deficiency of the LSDA 
by incorporating the Hubbard-like interaction term for $3d$ electrons. 
Formally, the LSDA + $U$ Hamiltonian is given by
\begin{eqnarray}
{\cal H}& &_{{\sf LSDA} + U} \nonumber \\ 
& & = {\cal H}_{\sf LSDA} \nonumber  -\left[ \frac{1}{2}UN(N-1)  
    - \frac{1}{2}J\sum_{\sigma} N^{\sigma}(N^{\sigma}-1) \right] \nonumber \\ 
& &~~ + \frac{1}{2}\sum_{\{m\},\sigma}V(mm';m''m''')n_{mm''}^{\sigma}
                              n_{m'm'''}^{-\sigma} \nonumber \\ 
& &~~ + \frac{1}{2}\sum_{\{m\},\sigma}
               \left[V(mm';m''m''')-V(mm';m'''m'')\right] \nonumber \\
& & ~~~~~~~~~~~~\times n_{mm''}^{\sigma} n_{m'm'''}^{\sigma}, \\ 
V& &(mm';m''m''') = \sum_{k=0}^{2l}c^{k}(lm,lm'')c^{k}(lm',lm''')F^{k}
\end{eqnarray}
where $n_{mm'}^{\sigma}$ is the $d$ occupation number matrix 
of spin $\sigma$,
$F^k$ is the Slater integral, and $c^k(l,m:l,m')$ is a Gaunt coefficient.
The conventional notations, $N^{\sigma} = \sum_{m} n_{mm}^{\sigma}$ 
and $N = \sum_{\sigma} N^{\sigma}$, are used.
Two main parameters in the LSDA + $U$ are the Coulomb $U$ and the 
exchange $J$ interactions.
These parameters are related to the Slater integrals 
by $U = F^{0}$ and $J = (F^{2}+F^{4})/14$. 
The ratio of $F^{4}/F^{2}$ is known to be constant 
around 0.625 for most $3d$ transition metal atoms\cite{Sawatzky}.  
We have used parameter values of $U$ = 6.0 eV and $J$ = 0.89 eV \cite{Uvalue}.

To determine the size of orbital magnetic moments, 
we have explicitly included the spin-orbit coupling 
in the Hamiltonian\cite{MacDonald}.  
In this way, our calculational scheme corresponds 
to a fully relativistic LSDA + $U$ method.
We have used 80 k-points sampling for the tetrahedron integration
in the irreducible Brillouin zone wedge, 
except for 40 k-points sampling in the LSDA + $U$ calculation for a supercell.
\begin{figure}[t] \label{fig1ps}
\epsfig{file=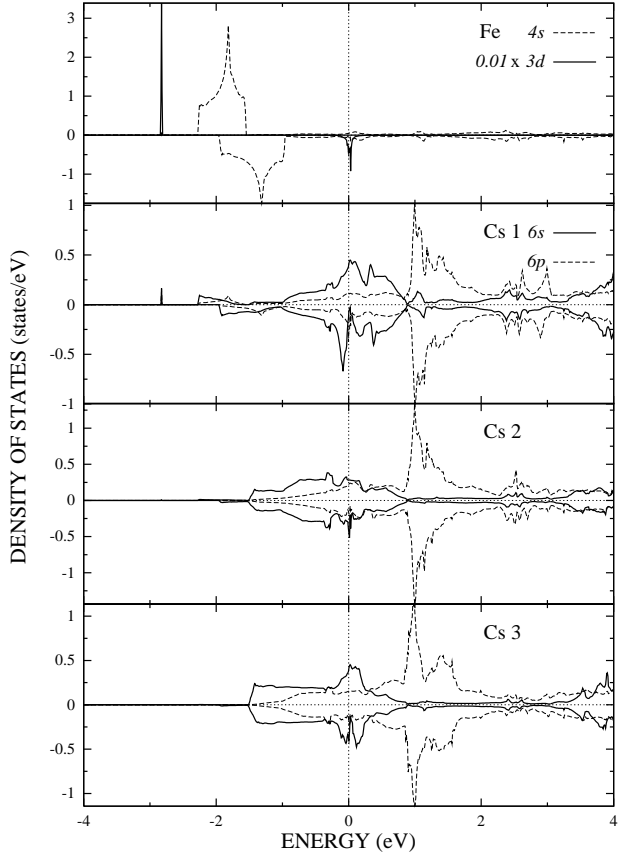,width=8.5cm}
\caption{ The LSDA DOS of 1Fe/5Cs ML structure.
Fe $3d$-bands are extremely localized 
and the shape of Fe $4s$ DOS exhibit a typical 2D DOS.
The spin polarization of Cs $6s$-bands is due to 
the hybridization with Fe bands. 
The spin magnetic moment of the first NN Cs is very small, $-0.09 \mu_B$.
}
\end{figure}
\begin{figure}[t] \label{fig2ps}
\epsfig{file=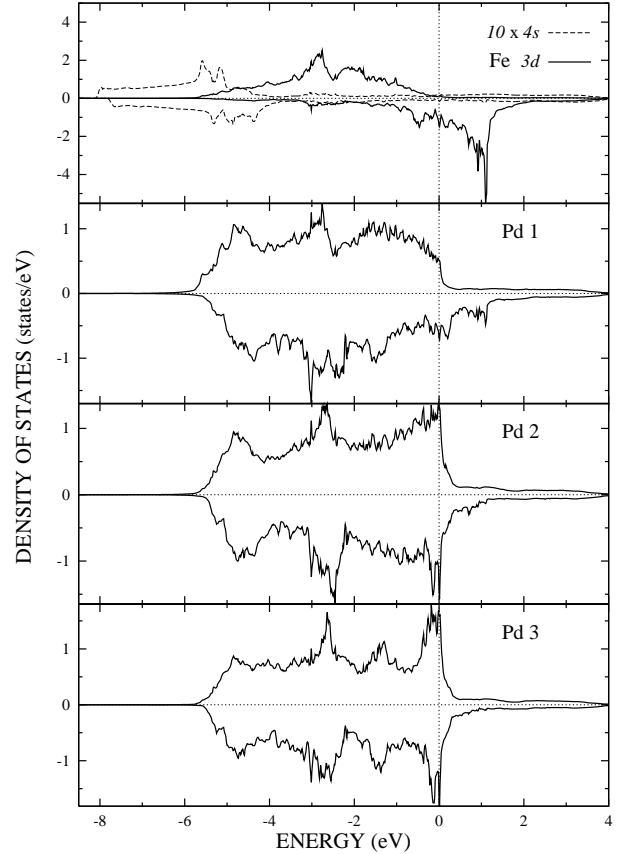,width=8.5cm}
\caption{ The LSDA DOS of 1Fe/5Pd ML structure.
The $4d$ DOS is shown for Pd.
The hybridization of Fe $3d$-bands with Pd $4d$ is large,
and so the orbital magnetic moment of Fe is negligible.
The spin magnetic moment of the first NN Pd is non-negligible, $0.30 \mu_B$.
}
\end{figure}

First, we have considered multilayered (ML) structures of Fe and Cs, 
which are stacks of the conventional bcc unit cells 
along the $\left[001\right]$ direction.
For 1Fe/3Cs structure
which contains a monolayer of Fe after three layers of Cs, 
the magnetic ordering is tested by the LSDA total energy calculations. 
The antiferromagnetic state is lower by only about 1 mRy
than the ferromagnetic state.
This result, however, is within the limit of the numerical precision, 
and thus one cannot ascertain which state is the ground state.  
Indeed, the experiment\cite{Beckmann} identifies the 
local magnetic moment formation, but the existence of the 
long range order is not confirmed. Other experimental tools 
are required to reveal the long range order in the system.
In the following discussion, 
we will consider only the ferromagnetic ordering state.
 
Figure 1 is the LSDA density of states (DOS) 
at each atomic site for 1Fe/5Cs ML structure.
The Fe-Fe bond lengths are $a = 6.05$ \AA~ in plane 
and $3a=18.15$ \AA~ out of plane.
The spin-up bands of Fe $3d$ is at about $-2.8$ eV below the Fermi level
$\rm E_F$. In fact, it is more appropriate to call Fe $3d$-levels 
rather than $3d$-bands.
The energy levels are well-defined outside of Fe $4s$- and Cs $6s$-bands 
so that the spin-up electrons of Fe $3d$ do not hybridize with other
electrons and form local levels. On the other hand,
the spin-down bands of Fe $3d$ are slightly mixed with Cs $6s$-bands 
at $\rm E_F$ and show an energy dispersion. 
Interestingly, the shape of Fe $4s$-bands is reminiscent of a typical 
2-dimensional (2D) DOS which has a logarithmic divergence in the band center 
and becomes constant at the band edges.
We have also found similar features for Fe ions in 1Fe/3Cs ML.
This suggests that the Fe $s$-$s$ interaction rapidly diminishes
over $2a = 12.10$ \AA, despite the delocalized characteristics
of $s$ electrons, and Fe ions are essentially isolated in the c-direction.

The spin splittings of Cs bands are seen for the first 
and the second nearest neighbor (NN) Cs of Fe,
and it becomes negligible for the third NN Cs.
The induced spin magnetic moments at Cs sites are 
$-0.09 \mu_B$ and $-0.02 \mu_B$ 
for the first and the second NN, respectively. 
The hybridization of Cs $6s$-bands with Fe bands, stronger in the
minority band, populates the spin-down states of Cs $6s$
and drives the spin magnetic moments at Cs sites to align 
antiferromagnetically to the spin of Fe ion. 
The spin and the orbital magnetic moments of Fe ion 
are, respectively, $\mu_S = 3.46 \mu_B$ and $\mu_L = 1.59 \mu_B$ in the LSDA.
When the LSDA + $U$ is applied, 
they become $\mu_S = 3.12 \mu_B$ and $\mu_L = 2.87 \mu_B$.
The Coulomb correlation effects is obvious, 
which enlarges the $3d$-level splitting
and increases the orbital magnetic moment.
The large orbital magnetic moment manifests that Fe $3d$ states are
localized in Cs.

A comparison with the system of Fe impurities in Pd 
will be helpful to understand Fe in Cs.
In Fig. 2, we show the LSDA DOS for 1Fe/5Pd ML structure,
of which the stacking unit is the face-centered-cubic (fcc) unit cell of Pd.   
We show for Pd sites only the $4d$ DOS.
The exchange splitting of Fe $3d$-bands is still large $\sim  3$ eV,
and the spin magnetic moment of Fe ion is $\mu_S = 3.12 \mu_B$ in Pd,
which are similar sizes to those of Fe in Cs.
However, the orbital magnetic moment of Fe ion 
is negligible $\mu_L = 0.09 \mu_B$ in Pd. 
The $3d$ orbitals of Fe are greatly deformed by the hybridization with Pd, 
and so Fe loses most of ionic nature in Pd. This feature is
very different from the case of Fe in Cs. 
The spin magnetic moments at the first and the second NN Pd sites 
are $0.30 \mu_B$ and $0.07 \mu_B$, respectively.
Total sum of the spin magnetic moment of Fe
and non-negligible spin magnetic moments of surrounding 
Pd atoms gives rise to the giant magnetic moment.

For a realistic simulation of Fe impurities in Cs films,  
we have constructed a supercell, 
the size of which is eight times larger than the bcc unit cell of Cs. 
The supercell has one Fe atom at the center
and fifteen Cs atoms at the right sites. 
Considering that the Fe-Fe interaction becomes
negligible over $2a=12.10 $ \AA, the size of the supercell is large enough 
to examine Fe impurity effects.
In the LSDA, the properties of Fe $3d$-bands are not much different from 
those of ML structures. 
The spin-up bands of Fe $3d$ shift down a little bit to higher 
binding energy (see Fig. 3). In the LSDA + $U$, notable differences are that
Fe $3d$ states disappear near $\rm E_F$ by the Coulomb correlation interaction, 
and the energy levels of Fe $3d$ are sharply split 
by the spin-orbit interaction in addition to the Coulomb correlation.
Hence, the spin-down Fe $3d$-bands no longer hybridize with Cs $6s$-bands.
In the inset of Fig. 3, the angular distribution of the occupied Fe $3d$
spin-down states are plotted, based on the orbital dependent occupancies
from the LSDA + $U$ calculation.
Although there is slight deviation from perfectly isolated states,
the angular distribution
is almost independent of the azimuthal angle rotation.
The symmetric angular distribution indicates that Fe $3d$ states are
well-localized in Cs.
Besides the localization of Fe $3d$ electrons, 
the 2D shape of Fe $4s$-bands in the ML structure is destroyed 
and Fe $4s$-bands become also localized in the supercell.
\begin{figure}[t] \label{fig3ps}
\epsfig{file=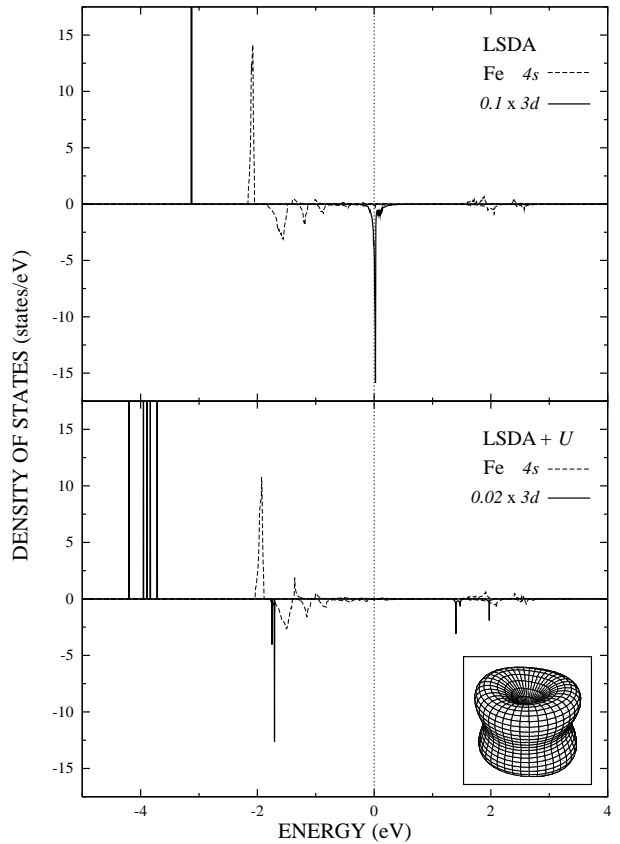,width=8.1cm}
\caption{ The LSDA (upper panel) and the LSDA + $U$ (lower panel) 
DOS's of Fe in the supercell structure. 
The magnitudes of the spin-up bands of Fe $3d$ are arbitrary. 
All the Fe bands, both of Fe $4s$ and $3d$, are localized.
In the LSDA, one notices a small energy dispersion of the spin-down bands 
of Fe $3d$ at $\rm E_F$ due to the hybridization with Cs $6s$-bands. 
In the LSDA + $U$, Fe $3d$-levels are split by the Coulomb correlation 
and the spin-orbit interactions.
Angular distribution of the occupied spin-down states 
of Fe $3d$ (inset) indicates that Fe $3d$ states are well-localized. 
}
\end{figure}
\begin{table}[b]
\caption{\label{cmm} The calculated magnetic moments ($\mu_B$) 
for the supercell structure. 
The LSDA + SO represents a relativistic LSDA calculation. 
In the LSDA + $U$, we have explicitly included the spin-orbit interaction.
The magnetic moments of Cs are sums of all the Cs moments in the supercell.
}
\begin{tabular}{ccc|cc|cccccc}
&Method      &&total $M$&&          &      & Fe   &&  Cs    & \\ \hline 
&LSDA        && 3.29    && $\mu_S$  &\vline& 3.59 &&$-0.29$ & \\ \hline
&LSDA + SO   && 4.62    && $\mu_S$  &\vline& 3.54 &&$-0.51$ & \\
&            &&         && $\mu_L$  &\vline& 1.60 &&$-0.01$ & \\ \hline
&LSDA + $U$~~&& 6.43    && $\mu_S$  &\vline& 3.14 &&$~~0.42$& \\
&            &&         && $\mu_L$  &\vline& 2.90 &&$-0.04$ &  
\end{tabular}
\end{table}

We summarize the calculated magnetic moments in Table~\ref{cmm}.
Because of charge transfer from Cs sites, 
Fe ion gets about 0.5 (0.8) more electrons in the LSDA (LSDA + $U$)
and these electrons go to the minority bands of Fe $3d$. 
While the transferred charge reduces the spin magnetic moment of Fe ion, 
the reduced spin magnetic moment is compensated by the enhanced orbital 
magnetic moment.
The spin magnetic moments of Cs are antiparallel (parallel) 
to the Fe spin in the LSDA (LSDA + $U$).
The direction of the spin polarization in Cs atoms can be understood 
by the band hybridization effects.
In the LSDA, the spin-down bands of Fe $3d$ hybridize with Cs $6s$-bands 
near $\rm E_F$ populating the spin-down states of Cs $6s$, 
whereas, in the LSDA + $U$, the occupied spin-down bands of Fe $3d$
are pushed down to about $-1.7$ eV below $\rm E_F$
and remain as local levels.
In the LSDA, the total magnetic moment of $M = 4.62 \mu_B$ is too small.
It is because the spin-orbit coupling in the LSDA is not so effective 
to polarize orbitals. 
The correct orbital ordering is obtained only with the strong 
Coulomb correlation interaction.
The total magnetic moment of $M = 6.43 \mu_B$ in the LSDA + $U$ 
is composed of Fe 94 \% and Cs 6 \% contributions~\cite{CoMM}. 
One can conclude from these results
that the electronic states of Fe ion are almost fully localized in Cs,
and the giant magnetic moment arises mainly from Fe ion itself rather than
the induced spin polarization of Cs atoms. 
In this sense, $3d$ electrons of Fe in Cs should be treated as
$4f$ electrons in rare-earth ions. It is desirable to check such a
large orbital moment in this system with experimental tools such as
the magnetic circular dichroism (MCD) or the resonant X-ray magnetic
scattering. 
The small discrepancy from the experiment ($\sim 7 \mu_B$) may be 
due to improper use of $g_J = 2$ in the analysis\cite{Beckmann}, 
which counts only the spin contribution.

In addition to the giant magnetic moments of Fe impurities in Cs films,
neither Fe coverage effects and nor moment change between Fe ions 
on the surface and in the bulk are observed\cite{Beckmann}.
In general, magnetic impurities on the surface of solids 
have larger magnetic moments than in the bulk.
On the surface, electrons are localized due to
lack of the degree of freedom in the surface normal direction.
These electrons, confined on the surface plane, results in
enhanced magnetic moments.
For Fe impurities in Cs films, Fe ions are fully localized
even in the bulk, and thus further localization of Fe electrons does 
not occur on the surface.
The negligible Fe coverage effects can be understood considering 
the short interaction range ($2a=12.10 $ \AA) between Fe ions.
The magnetic moment rarely depends upon the Fe concentration 
on the surface of Cs films, once excluding effects of
the lattice relaxation and the rearrangement of surface atoms.

In conclusion, we have found that Fe impurities 
in Cs films have a different behavior from those in Pd films,
and that the giant magnetic moments of Fe impurities in Cs
originate mainly from the localized Fe ions but not from the spin 
polarizations of the neighboring Cs atoms.
The contribution from the orbital magnetic moment of Fe $3d$ electrons
is comparable to that
from the spin magnetic moment, as in the $4f$ rare-earth ions. 
The Coulomb correlation interaction at the Fe site should be properly 
considered to get consistent results with experimental data.

Acknowledgements$-$
We are thankful to S. J. Youn, J. H. Park, and J.-H. Cho for 
helpful discussions. 
This work was supported by the KOSEF (1999-2-114-002-5)
and in part by the Korean MOST-FORT fund.
Computations were done at the POSTECH Supercomputing Center.


\begin{thebibliography}{99}
\bibitem{Uvalue} The results are not sensitive to $U$ value:  
         with $U$ = 4.0 eV, essentially the same results are obtained.
\bibitem{CoMM} For Co ion in the supercell structure, the total magnetic
         moment of $M = 5.81 \mu_B$ is obtained in the LSDA + $U$.
         In this case, Cs contribution to the total magnetic moment is
         slightly enhanced to about 13 \%.
\end{thebibliography}
\end{document}